\documentstyle[aps,preprint,eqsecnum]{revtex}
\newcommand{\al}{\alpha}
\newcommand{\bt}{\beta}
\newcommand{\rt}{\right}
\newcommand{\lt}{\left}
\newcommand{\m}{\mu}
\newcommand{\n}{\nu}
\newcommand{\ga}{\gamma}
\newcommand{\la}{\lambda}
\newcommand{\de}{\delta}
\newcommand{\pa}{\partial}
\newcommand{\ka}{\kappa}
\newcommand{\th}{\theta}
\newcommand{\ze}{\zeta}
\newcommand{\pr}{\prime}
\newcommand{\f}{\frac}
\newcommand{\r}{\bar}
\newcommand{\be}{\begin{equation}}
\newcommand{\ee}{\end{equation}}
\newcommand{\se}{\section}
\newcommand{\sse}{\subsection}
\newcommand{\bey}{\begin{eqnarray}}
\newcommand{\eey}{\end{eqnarray}}
\newcommand{\bi}{\bibitem}
\newcommand{\ct}{\cite}
\begin{document}

\title{SUPERSPACE FORMULATION OF YANG- MILLS THEORY II: INCLUSION OF
GAUGE INVARIANT OPERATORS AND SCALARS}

\vspace{.7in}

\author{Satish D. Joglekar \footnote{e-mail address :
sdj@iitk.ernet.in}  and  Bhabani Prasad Mandal\footnote{
e-mail address : bpm@iitk.ernet.in}}

\vspace{.6in}

\address{Department of Physics \\ Indian Institute of
Technology,Kanpur\\Kanpur 208016 (INDIA)}
\maketitle

\vspace{20mm}

\begin {abstract}
In a superspace formulation of Yang-Mills theory previously
proposed, we show how gauge-invariant operators and scalars
can be incorporated keeping intact the (broken) $Osp(3,1|2)$
symmetry of the superspace action. We show in both cases, that
the WT identities can be cast in a simple form
$\f{\pa\bar{W}}{\pa\th}=0$.
\end{abstract}

\newpage

  \se{INTRODUCTION}
     The theory of Yang-Mills fields forms the backbone of all the
     successful high energy physics of today  viz. the Standard model.
     Yang-Mills fields are characterized by a nonabelian local gauge
     invariance.Consequence of gauge invariance are formulated as the
     Ward-Takahashi (WT) identities of gauge theories. A study of these
     is highly important in any dealing with the gauge theories. In
     particular, establishment of renormalizability requires use of these.
     The derivation and study of WT identities and of the
     renormalization
     program  in gauge theories is facilitated greatly by the introduction
     of the alternate global symmetry of the effective action, viz. the
     BRS (Becchi- Rouet -Stora) symmetry, \ct{brs,zinn,lee}.
     Hence any program that
     will shed light  on BRS symmetry, Ward identities, renormalization
     of gauge theories etc. is of value.

     BRS transformations contain an anti-commuting parameter. This
     has naturally lead to construction of superfield /superspace
     formulation
     of gauge theories in order to make the underlying BRS structure
     evident in a simple way \ct{ss1,ss2}.

     With a view to simplify the WT identities and renormalization program
     in gauge theories, (especially for gauge invariant operators) an
     improved superspace formulation was proposed \ct{sdj}. This
     formulation
     has the advantage that the superfield were completely unrestricted
     ( and not constructed  by hand as in earlier formulations \ct{ss1,ss2}
     ). It
     further had an $ Osp(3,1|2) $ broken invariance,which unlike earlier
     formulation must not a very formal device \ct{ss2}, but
     one where superspace
     rotation could be actually carried out.
     Further, the source terms for composite BRS variation operators,
     so
     crucial to a simple formulation of WT identities were generated from
     within. The sources for BRS variations and fields came from
     supermultiplets of (super)sources.These led to a simple formulation
     of WT identities \ct{sdj1}, in the elegant form
     $ \f{\pa\bar{W}}{\pa\th} = 0 $.
     These identities were further understood as arizing from a broken
     $ Osp(3,1|2) $ symmetry of the superspace action \ct{osp}. The
     formulation of ref.\ct{sdj} has also lead to an understanding of
     interrelation between renormalization transformations in gauge
     theories \ct{ren}.The formulation of ref.\ct{sdj} has also been
     generalized to incorporate the most general BRS/anti-BRS symmetry
     in linear gauges \ct{anti}.

     A difficult problem in the renormalization program of gauge
     theories is the problem of renormalization of gauge invariant
     operators. This problem is relevant in phenomenological applications
     of QCD through the use of operator product expansion \ct{gross}.
     Though solved long ago \ct{ann}, the long treatment could well be
     simplified. We expect the present superspace formulation to do
     this, because it simplifies the WT identities in form, and hence
     hopefully their solution too. First step towards solution to this
     problem in this formulation is to show how to incorporate
     the gauge invariant operators in the formulation without spoiling
     $ Osp(3,1|2) $ symmetry (broken). Next step is to show that the WT
     identities retain their simple form in presence of gauge invariant
     sources. We propose to do these two steps in the present work.
     The
     final step of solving the WT identity will be reported elsewhere.

     In passing, we also show that the scalars can be incorporated in a
     straightforward manner in this formulation. This will have application
     in the superspace formulation of spontaneously broken gauge theories.

     We now summerize the plan of the paper. In section II, we
     introduce the superspace formulation of ref.\ct{sdj} and
     propose its extension to include gauge-invariant sources
     without spoiling broken $ Osp(3,1|2)$ symmetry of the Lagrange
     density. In section III, we show that the proposed generating
     functional does in fact contain all the informations about the
     Green's functions with one insertion of a gauge invariant
     operator. In section IV, we exhibit how scalars in arbitrary
     representation of the gauge group  can be included
     straightforwardly. In section V, we show that in both cases the
     WT identities retain their simple form $ \f{\pa\r{W}}{\pa\th}
     = 0 $ . Appendices A and B deal with general structure of
     gauge -invariant operators in Minkowski space and
     superspace. In appendix C we have listed the Jacobi identities
     and commutation/anti-commutation relations satisfied by the
     superspace covariant derivatives.

\se{SUPERSPACE ACTION TO INCLUDE GAUGE-INVARIANT OPERATOR
INSERTION}
\sse{  Superspace notations and superspace action}

 In this section we shall briefly introduce the notation [ For more
 details, the reader is referred to ref.\ct{sdj}]. We shall then
 introduce  the superspace action for ordinary gauge theories and
 show how it is to be generalized to include an insertion of a
 gauge invariant operator of arbitrary dimensions and Lorentz
 structure.

 We shall work in a 6-dimensional superspace $ \bar{x}^i =
 (x^{\m},\la,\th) $ with two anti-commuting coordinates $\la$ and
 $\th $.The superspace  formulation of ref.\ct{sdj} utilizes an
 anti-commuting antighost superfield $\ze^\al (\bar{x})$ and a
 commuting superfield $\bar{ A}^{\al}_i (\bar{x}).$ The former is a
 scalar under rotations \ct{osp} characterized by the group $
 Osp(3,1|2)$ and the latter is a covariant vector.The sources for
 fields and composite operators for the BRS variations also come
 in the form of a (commuting) scalar superfields $t^\al (\bar{x})$
 and an (anticommuting) vector superfield $\bar{K}^i(\bar{x})$
 . We will not go into further details here for which reader is
 referred to references \ct{sdj} and \ct{osp}.We only state the
 superspace action :

 \bey
 \bar{S}&=& \int d^4x [ -\frac{1}{4} g^{ik}g^{jl} \bar{F}_{ij}
 \bar{F}_{kl}] + \int d^4x \frac{\pa}{\pa\th}\lt\{\bar{K}^i(\bar{x})
 \bar{A}_i(\bar{x}) + \ze^\al (\bar{x}) [ \pa^\m A^\al_\m (\bar{x})
 +\frac{1}{2\eta_0}\ze_{,\th}^\al +t^\al]\rt\} \nonumber \\
 &\equiv & \bar{S_0} +\bar{S_1} \nonumber \\
 &\equiv & \int d^4x \bar{\cal{L}}_0 +\int d^4x \bar{\cal{L}}_1
 \eey
where $\bar{F_{ij}}$ is the superspace generalization of the field
strength \ct{sdj} [The index $ i$ runs over 0 to 5 while the
index $\m$ runs over 0 to 3.]

The superspace generating functional for the superspace action is
given by

\be
\bar{W}[\bar{K}(\bar{x}),t(\bar{x})] = \int \{dA\}\{d\ze\}\,\exp(i
\bar{S}[\bar{A}
,\ze,K,t])
\ee
where the measure has been defined in ref.\ct{sdj}.

$\bar{\cal{L}}_0 $ is invariant under $Osp(3,1|2)$ superspace
rotations. $ \r{W} $ is related to the generating functional of
ordinary Yang-Mills Green's functions by \ct{sdj}

\be
\int [dk^4][dk^4_{,\th}] \r{W} [\r{K},t] = W \lt[ K^\m_{,\th}
(\r{x}),K^5_{,\th} (\r{x}),-t_{,\th} (\r{x}),K^\m (\r{x}),K^5
(\r{x}),t(\r{x}) \rt]
\ee

\sse{ Inclusion of gauge-Invariant operators}

We want to generalize $\bar{\cal{L}}_0$ to include source term for
a gauge invariant operator so that it still has $Osp(3,1|2)$
invariance.

Now consider a typical gauge invariant operator $ O = O_{\m_1
\cdots \m_n}[A]$. Firstly  $O$ belongs to a representation of the
Lorentz group $ O(3,1)$ and not to a representation of
$ Osp(3,1|2)$. Secondly, it only contain the $ x-$dependent
part of the gauge fields $ A_{\m}(\bar{x}) $ viz. $A_{\m}(x)$
(and its space time derivatives). Now in order to construct a
superspace analogue of $O$ viz. $\bar{O}$, we have to (i) alter
the external index structure (ii)change $A_\m (x)$ to superfield
$\r{A}_i (\r{x})$ (iii) change space-time derivatives to
superspace derivatives. This will be done as follows,

As shown in appendix A, a gauge-invariant operator can be written
entirely in terms of $ D^{\al\bt}_\m $ and $ F^{\al}_{\n \la}$'s.
We replace each $D^{\al\bt}_\m $ by the superspace covariant
derivative operator $ D^{\al\bt}_i = -\de^{\al\bt}\vec{\pa_i} +
gf^{\al\bt\ga}\r{A}^\ga_i(\r{x}) $ and each $ F^{\al}_{\n\la}$
by $ \r{F}^\al_{jk} (\r{x})$  defined in \ct{sdj}.
The contracted indices continue to be contracted but now range over
$0$ to $5$ instead of $0$ to $3$. One, thus, obtains an operator
$\r{O}_{i_1 \cdots i_n}[\r{A}(\r{x})]$ which belongs to a
representation of $Osp(3,1|2)$. We now introduce a superspace
source $ \r{N}^{i_1\cdots i_n}(\r{x})$, assumes also belong to an
 appropriate representation of $Osp(3,1|2)$ such that $
 \r{N}^{i_1\cdots i_n}(\r{x}) O_{i_1\cdots i_n}(\r{A}) $ is an
 $Osp(3,1|2)$ scalar. We then modify ${\cal{L}}_0$ to
a
 \be
 \r{{\cal{L}}^\pr_0} = \r{{\cal{L}}_0}+ \r{N}^{i_1\cdots i_n}(\r{x})
 \r{O}_{i_1\cdots i_n}[\r{A}]
 \ee
 and define the generating functional $ \r{W}^\pr $ accordingly
 so that

 \be
\r{W}^\pr[\r{K},t,\r{N}] = \int \{dA\}\{d\ze\}\, \exp(i\int
 d^4x[{\r{\cal{L}}^\pr_0} +{\cal{L}}_1])
 \ee
Then $\lt.{\f{\de\r{W}^\pr}{\de \r{N}^{i_1\cdots i_n}}}\rt|_{\r{N}=0}$
generates the Green's functions of the superspace theory with one
insertion of $\r{O}_{i_1 \cdots i_n} $.
Among these we are interested in those with one insertion of
$\r{O}_{\m_1\cdots
\m_n}$. We  shall show how we can recover the Green's functions of
the Y-M theory with one insertion of $ O_{\m_1 \cdots \m_n}$ in a
later section.[ See Eqn.3.9]

\se{ EVALUATION OF $ \r{W}^\pr \lt[\r{K}, t,\r{N} \rt]$}

 We shall show, in this section, how the generating functional
 $ \r{W}^\pr $ can be evaluated partially, by performing integrals
 over $ A_{i,\th} $ and $ A_{i,\la}$. The procedure is very much
 the same as that in ref.\ct{sdj}. Hence, we shall be brief and
 indicate only main steps successively.

 (1.) In


 \be
 \r{W}^\pr = \int \{dA\}\{d\ze\}\,\exp\lt[i\r{S} + i\int d^4x
 \r{N}\r{O}\rt]
 \ee
 dependence of $ \r{S}+\int\r{N}\r{O} d^4x $ on $A_{i,\th}$ and
 $A_{i,\la}$ arising {\em solely } out of dependence of $ \r{A}
 (\r{x}) $ on these  via


 \be
 \r{A}_i = A_i (x) + \th\tilde{A}_{i,\th} +\la\tilde{A}_{i,\la}+
 \la\th\tilde{A}_{i,\la\th}
 \ee
  is immaterial as if can be removed away by equation of motion of
  $ \r{A} $ [This is explained in detail in ref.\ct{sdj}. Hence
  only the bona fied dependence of $ \r{S} + \int\r{N}\r{O} d^4x $
  on $ A_{i,\th}$ and $ A_{i,\la} $ need to be retained.

  (2.) As the original $ \r{W} $ does not depend on $ A_{i,\la\th}$
  and as these are not dynamical variables, we shall omit ( put to
  zero ) terms in $ \r{N}\r{O} $  depending on $ A_{i,\la\th} $ in
  evaluating $ \r{W}^\pr $.

  (3.) In $ \r{W}^\pr $ we are only interested in the terms to the
  first order in $\r{N} $. Hence we consider
 \begin{eqnarray}
 \bar{W}^\pr = && \int \{dA\}\{d\zeta\} \exp{\lt(i\bar{S}
 \rt)}\lt[1+i\int d^4x
 \bar{N}\bar{O}\rt] \nonumber \\
 \equiv && \bar{W} + i \ll \int d^4x \bar{N}\bar{O} \gg
 \end{eqnarray}
and evaluate it in this form. Then later we put it back in the form
of exponential.

(4.) In evaluating the term $\ll \r{N}\r{O} \gg $ in (3.3), we can
use the equation of motion for $ \r{S} $ ( ie. in absence of
$\r{N}\r{O} $ ) in it as this term is already of order $\r{N} $.

(5.) In performing the integrals over $ A_{\m,\la},
A_{\m,\th},c_{4,\la}, c_{5,\th}, c_{5,\la} $, we can choose instead
as integration variables $ F_{\m 5},F_{\m 4},F_{44},F_{55},F_{45} $
as the Jacobian of the relevant transformations is one. [ See
relation of Eq.(3.3) of ref.\ct{sdj} for elaboration]

(6.) The equation of motion for $F_{\m 5},F_{\m 4},F_{44}, $ and $F_{55}$
in absence of $\r{N}\r{O} $ term read
\bey
F_{4\m} +K_\m - \pa_\m\ze &=& 0 \nonumber \\
F_{5\m} &=& 0 \nonumber \\
F_{55} &=& 0 \nonumber \\
F_{44}-2K^5 &=& 0
\eey

(7.) In relating $\r{W}^\pr $ to the Y-M generating functional with
one insertion of $ O $, we shall be needing

\be
W^\pr =\int \lt[dK^4 \rt]\lt[dK^4_\th\rt] \r{W}^\pr
\ee
just the same way as equation (2.3) involves $ \int
\lt[dK^4\rt]\lt[dK^4_\th\rt] \r{W} $. Hence we shall begin with
(3.5) which in effect puts to zero $ c_4 $ and $c_{4,\th}$. Then
the modified equation of motion  for $ c_{5,\la} $ or equivalently
$ \tilde{F_{45}} = F_{45}\lt. \rt|_{c_{4}=0,c_{4,\th}=0} $ reads

$$ F_{45} = 0 \eqno(3.4a) $$

(8.) Now, it is easy to show that the effect of doing integration
over $ A_{\m,\th}, A_{\m,\la},$ $ c_{4,\la},c_{5,\th},c_{5,\la} $ in
(3.3) is simply to use equation of motion  (3.4) and (3.4a) in the
$\r{N}\r{O} $ term. [Hence, we assume dimensional regularization
which puts to zero $\de^4(0) $ and derivatives of $ \de^4(x) $ at
$ x=0$ ]. Thus, we obtain

\be
\int\lt[dK^4\rt]\lt[dK^4_{,\th}\rt] \r{W}^\pr\lt[\r{K},t,\r{N}\rt] \simeq
\int \lt[dK^4\rt]\lt[dK^4_{,\th}\rt] \lt[\r{W} +i\ll \int d^4x
\r{N}\r{O}^\pr \gg \rt]
\ee
where $\r{O}^\pr $ is obtain from $ \r{O} $ by following operations

(a) Put $ c_4 $ and $ c_{4,\th} $  equal to zero. \\
(b) Remove  $ A_{i,\th} $ and $ A_{i,\la}$ dependence of $\r{O} $
arising solely out of $ \r{A}(\r{x})$ \\
(c) Put $ A_{i,\la\th} $ to zero. \\
(d) Express $\r{O}$ in terms of $ F_{\m 5},F_{\m 4},F_{44},F_{55},F_{45}
$ instead of $ A_{\m,\th}, A_{\m.\la} \cdots c_{5,\la} $; and use
(3.4) and (3.4a) in them. In particular note that  it replaces
$ F_{5 \m}, F_{55},F_{45} $ to zero and $ F_{4 \m} $ by $ -(K_\m-\de_\m\ze)
   $ and $ F_{44}$ by $ 2K^5 $ .\\
(9.) We shall show in the appendix B that $ \f{\de \r{O}_{\m_1\cdots
\m_n}}{\de F_{4\m}} $ and $\f{\de \r{O}_{\m_1\cdots \m_n}}{\de F_{44}} $
contain terms of which contains $ F_{5 \m}, F_{55} $, or $ F_{45}$
as a factor, the latter being zero by equation of (3.4). Hence the
statement made in 8(d) above {\em all} dependence of $ O$ on $
A_{\m,\th},A_{\m,\la}\cdots ,c_{5,\la} $ and of course $ c_4 $ and
$c_{4,\th} $  drops out. This leaves a possible $ c_5 $
dependence  alone while is impossible as $ c_5 $ carries a ghost
number one while $ \r{O}$ does not and there is no field left in $ \r{O}
^\pr $ to compensate the ghost number of $ c_5 $. Hence
\be
\r{O}^\pr_{\m_1 \cdots \m_n} = O_{\m_1 \cdots \m_n}
\ee
Note that above argument does not generally apply to $ \r{O}_{i_1
\cdots i_n} $ with $ i_k = 4 $ or 5 for some k. This should not
bother us as we are {\em not} interested in insertion of such
spurious operators. They have been added to recover formal
$ Osp(3,1|2) $ invariance of $\r{{\cal L }}_0^\pr $  which is
expected to be useful in formal manipulation of $ \r{W}^\pr $.
Hence we note that
\bey
W^\pr &=&
\int\lt[dK^4\rt]\lt[dK^4_{,\th}\rt]\,\r{W}^\pr\lt[\r{K},t,\r{N}\rt]
\nonumber \\
&=& \int \lt[dK^4\rt]\lt[dK^4_{,\th}\rt]\exp{\lt(i\r{S} +i\int d^4x
\,\r{N}_{\m_i \cdots \m_n} O_{\m_i \cdots \m_n}[A]\rt)} +
0\lt(\hat{N}_{i_1\cdots i_k \cdots i_n }\rt) + 0(N^2)
\eey
where $ \hat{N}_{i_1\cdots i_k\cdots i_n} $ denotes a source with
at least one $ i_k = $ 4 or 5 .

(10.) Now the integral on the right hand side of (3.8) can be done
as in ref.\ct{sdj}. The term $\r{N} O $ does not interfere in any way
with $ A_{\m,\th},A_{\m,\la},\cdots c_{5,\la} $ integrations as it
does not depend on anything but $A_\m(x)$. The result is self
evidently,
\bey
W^\pr = \int \lt[dA\rt]\lt[dc_5\rt]\lt[d\ze\rt]\exp\lt(\rt. i\lt[\rt.\int
{\cal L}_0 \,d^4x &+& \int \r{N}_{\m_1\cdots \m_n }O_{\m_1\cdots
\m_n} \,d^4x \nonumber \\
&+&\mbox{source terms for fields and BRS composite
operators}\lt.\rt]\lt.\rt) \nonumber \\
= W\lt[K^{\al\m}_{,\th},K^{\al 5}_{,\th},-t^\al_{,\th},K^{\al\m}
(\r{x}), K^{ \alpha 5} (\bar{x}) ,t^\al (\bar{x}) ,\r{N}\rt]&.&
\eey
where $ W $ on the right hand side of Eq.(3.9) is $ W $  Eq.(2.3)
modified to include gauge-invariant source $\int\r{N} O \,d^4x $ .
Equation (3.8) together with Eq.(3.9) shows the equivalence of
$ W^\pr = \int \lt[dK^4\rt]\lt[d^4K^4_{,\th}\rt]\,
\r{W}^\pr\lt[\r{K},t,\r{N}\rt] $ and the Yang-Mills generating
functional with one insertion  of $ O_{\m_1\cdots\m_n} $ viz $ W $
of (3.9) upto terms irrelevant for our purpose.

We have thus, in Eq.(2.5) constructed a generating functional $
\r{W}^\pr $, contains $\r{{\cal L}}^\pr_0 $ (see Eq.(2.4)) that
possesses symmetry, which yields us correctly the Green's function
of Yang-Mills theory with {\cal one} insertion of an arbitrary
gauge-invariant operator $ O_{\m_1\cdots \m_n} $. We shall consider
the WT identity satisfied by $ W^\pr $ in see.V.

\se{INCLUSION OF SCALARS}

If one is to apply this superspace formulation to the discussion of
Higgs Mechanism say in the Weinberg Salam model, one must show how
the scalar fields can be incorporated in superspace formulation. In
this section we shall show that the scalars can be incorporated
trivially in the superspace formulation and show in sec.V that the
corresponding WT identities continue to retain the simple form
$ \f{\pa\r{W}}{\pa\th} =0 $

Consider a set of scalars in some representation of gauge group
$ G $. Let $ \{T^\al\}$ be the representation of generators of $G$
corresponding to the representation to which the scalars belong. Let
the scalars be represented by a real column vector $ \Phi(x)$. The
covariant derivative of $\Phi$ is
\be
D_\m\Phi =\lt(\pa_\m -igT^\al A^\al_\m \rt)\,\Phi
\ee
We now introduce a scalar superfield ( a group multiplet )
$\r{\Phi}(\r{x}) $ which transforms as a scalar under $ Osp(3,1|2)$
and has as its first component the column vector $\Phi(\r{x})$; ie.

\be
\r{\Phi}(\r{x}) = \Phi(x) +\th\tilde{\Phi}_{,\th}
+\la\tilde{\Phi}_{,\la} +\la\th\tilde{\Phi}_{,\la\th}
\ee
 Let the usual Yang-Mills action including scalars be
\be
 {{\cal L}}_{0\phi} = -\f{1}{4} F_{\m\n}F^{\m\n} +
 \f{1}{2} {\lt(D_\m\Phi\rt)}^T \lt(D^\m\Phi\rt) +V\lt(\Phi\rt)
 \ee
 where $ V\lt(\Phi\rt) $ contains the mass terms and interactions
 (independent of $\pa_\m\Phi $). We then define the generating
 functional for the new superspace action by
\be
\r{W}_\phi \lt[\r{K},t ,J(\r{x})\rt] = \int \lt\{dA\rt\}\lt\{d\ze\rt\}
\lt\{d\Phi\rt\}\,\exp{i\r{S}_\phi\lt[ A, \ze , \Phi ,K ,t,J \rt]}
\ee
where
\be
\r{S}_\phi\lt[ A,\ze , \Phi ,K, t,J\rt] =\r{S} + \int d^4x
\f{\pa}{\pa\th}\lt[ {J(\r{x})}^T \Phi(\r{x})\rt] +\int d^4x
\lt[\f{1}{2}{\lt[D_i\Phi(\r{x})\rt]}^T\lt[D_j\Phi(\r{x})\rt]
g^{ij}+ V \lt(\Phi(\r{x})\rt)\rt]
\ee
and
\be
\lt\{d\Phi(\r{x})\rt\} \equiv \prod _i\lt\{d\Phi_i(\r{x})\rt\}
=\prod _i\prod _x d\Phi(x)\, d\Phi_{i,\la}(x) \, d\Phi_{i,\th}(x)
\ee

The $\r{W}_\phi$ of Eq.(4.4) is nothing but a straightforward
generalization of $\r{W}$ of Eq.(2.2). We shall now show that
$\r{W}_\phi $ contain all informations about the Green's function
$S$  of the Lagrange density of Eq.(4.3). Incidentally note that
the $\r{{\cal L}}_{0\phi}$ [with obvious interpretation] has the
full symmetry under superspace rotations of $ Osp(3,1|2) $.

To do this, we need only to concentrate on two new integrals over
$\Phi _{i,\la} (x)$ and $
\Phi_{i,\th}(x) $. This is done in a few steps.

(i) Firstly dependence of $\r{S} $ on $\Phi_{i,\la}$ and
$\Phi_{i,\th}$ arising solely from that of $ \Phi_i$ on these
variables via. Eq.(4.2) can be ignored completely on account  of
equations of motion of $\Phi_{i}$ \ct{sdj}.\\
(ii) Secondly $\Phi_{i,\la\th} $ do not appear in $\r{W}_\phi $ as
dynamical variables.\\
(iii) Consider now the term in $\r{S}$ dependent on $\Phi_{i,\la}$
and $\Phi_{i,\th}$. They are (with corresponding integrals )
\be
\int \lt[d\Phi_{,\la}\rt]\lt[d\Phi_{,\th}\rt]\,\exp{\lt(i\int d^4x\lt\{
J^T_{,\th}(\r{x})\Phi(x) -J^T(x)\Phi_{,\th}(x)+{\lt(\Phi_{,\la}-igT^\al
c^\al_4\Phi\rt)}^T\lt(\Phi_{,\th}-igT^\al c^\al_5\Phi\rt)\rt\}\rt)}
\ee

We can now change the variables of integration to
${\lt(D_4\Phi\rt)}_i $ and $ {\lt(D_5\Phi\rt)}_j $
themselves. The Jacobian of the transformation is one .Integration
over $ {\lt(D_4\Phi\rt)}_i$ yields on anti-commuting $\de$-function
$\displaystyle \prod_{i,x}{\lt(D_5\Phi\rt)}_i(x)$
. This can be used to replace
$ \Phi_{,\th} $ in the source term by
$\lt(-igT^\al c^\al_5\Phi\rt) $. Then ${\lt(D_5\Phi\rt)}_i$ integral
is done trivially. Thus, the net scalar dependent piece in the
expression for $\r{W}_\phi $ read
\be
\int \lt[d\Phi(x)\rt] \, \exp{\lt(i\int d^4x \lt[{\cal L }_{0\phi}
+J^T_{,\th}(\r{x})\Phi(x) -J^T(\r{x}) ig T^\al c_5^\al \Phi(x)
\rt]\rt)}
\ee
The last term in the exponent of (4.8) gives the term for BRS
variation of $\Phi $ :
\be
\de\Phi_{BRS} = -igT^\al c^\al_5 \Phi \de\Lambda
\ee
The rest of the integrals can be carried out straightforwardly
as in ref.\ct{sdj}. The result is
\be
\r{W}_\phi \lt[ \r{K},t,J(\r{x})\rt] = \prod_{\al
,x}\lt[\de\lt(K^{\al 4}(x)\rt)\de (K^{\al 4}_{,\th}(x))\rt]\,
W\lt[K^\m_{,\th},K^5_{,\th},-t_{,\th},K^\m,K^5,t,J ,J_{,\th}\rt]
\ee
where $ W$ is the generating functional for the  action ${\cal
L}_{0\phi}$  of Eq.(4.3) with sources for BRS variations of $
A_\m\,,c\,,\r{c} $ and $\Phi $ included.
This concludes the discussion of how scalars can be included in
the present superspace action.

\se{ WT IDENTITIES }

In this section, we shall give a brief derivation of the WT
identities of the gauge theories with (i) one insertion of gauge
invariant operator  (ii) scalars. The derivation is much the same
way as in ref.\ct{sdj1}. Hence we only explain the steps qualitatively.

The steps are :
(i) Evaluate $\f{\de\r{W}}{\de\th} $ from the integrated out form.\\
(ii) Recognize that these are  just the terms in the BRS variation
 of the integrated out action under the standard BRS
 transformations of the fields \ct{sdj1}

 We only mention new relevant points.

 \sse{Gauge invariant operator insertion}
 We call
 \be
 W^\pr = \int\lt[d^4K\rt]\lt[d^4K_\th\rt]\,\r{W}^\pr\lt[\r{K},t,\r{N} \rt]
 \ee
 In $ W^\pr $, the exponent now contain a new term $ \r{N}_{\m_1\cdots
 \m_n }(\r{x})\,O_{\m_1\cdots \m_n}[A] $. It will contribute to
 $\f{\pa W}{\pa\th} $ only through $\f{\pa \r{N}}{\pa \th}$. But we can
 always set $\f{\pa \r{N}}{\pa\th} =0 $ in the end being an independent
 field of $x^\m$. In the BRS variation, the new term does not
 contribute as $ O[A] $ is a gauge invariant  operator. Hence
 the proof of WT identities goes through as before, and we have
 \be
 \f{\pa W^\pr}{\pa\th} =0
 \ee
 $\lt[\mbox{Omitting } 0\lt(\hat{N}_{i_1\cdots i_k\cdots i_n},\,0(N^2),\,
  0\lt(\f{dN}{d\th}\rt) \mbox{terms } \rt) \rt]$

\sse{Scalars}

In this case there are the following new terms in the integrated
out action :
\be
\int d^4x \lt\{ J^T_{,\th} (\r{x})\Phi(x) -J^T(\r{x}) ig T^\al
A^\al_\m \Phi + \f{1}{2} {\lt(D_\m\Phi\rt)}^T\lt(D_\m\Phi\rt)
-V[\Phi] \rt\}
\ee
 In taking $\f{\pa\r{W}}{\pa\th} ,\,\,\, -J^T(\r{x}) ig T^\al A^\al_\m\Phi
 $ terms contributes. There is no further contribution. But this
 new contribution is recognized as exactly being the BRS variation
 of $J^T_{,\th}\Phi(x) $, other new term being BRS invariant.
 Hence the derivation of \ct{sdj1} goes through leading to the
 result
 \be
 \f{\pa\r{W}}{\pa\th} = 0
 \ee
 Thus the WT identities continue to retain their elegant form.
\newpage

\appendix
\se{General structure of a local gauge invariant operator}

We shall derive a very simple result in this appendix. It is about
the general structure of a gauge-invariant operator constructed out
of Yang-Mills fields only. The result, while it is rather trivial,
 is very much needed in the discussion of superspace
 generalization.

  We shall show that for a simple gauge group G, the most general
  local gauge invariant operator constructed out of Yang-Mills
  fields has the structure
  \be
  O =\sum \, t^{\al_1\cdots \al_n}\,X^{\al_1}_{(1)}\cdots \, X^{\al_n}_{(n)}
  \ee
  Where each of the $ X^{\al_k}_{(k)} $ can be written as a string
  of covariant derivatives, $D^{\al\bt}_\m $ acting on one field
  strength, $  F^\ga_{\n\la} $. For example,
  \be
  X^{\al_1}_{(1)} = D^{\al_1\bt}_\m \,D^{\bt\ga}_\n\,
  F^\ga_{\la\de}
  \ee
  The $ X_{(k)}^{\al_k}$ 's can be distinct for each $k$. Further $
  t^{\al_1\cdots \al_n} $ is a group tensor. $\Sigma $ denotes
  formally a sum of terms of this form.
  That $O $ of $(A1)$ is a gauge invariant operator is obvious.
  This is so since  each $ X^{\al_k}_{(k)} $ is a group vector and
  $ t^{\al_1\cdots\al_n}$ is a group tensor.

  To prove $ (A1) $, imagine expanding an arbitrary local
  gauge-invariant operator $ O $ of dimension $ D $ as
  \be
   O =\sum_{p=0}^D \, g^p\, O_{(p)}\lt[A\rt]
   \ee

   As, $ O $ is invariant under
   \be
   \de A^\al_\m =-\pa_\m\th^\al(x) + g f^{\al\bt\ga} A^\bt_\m
   \th^\ga (x)
   \ee

   $ O_{(0)}\lt[A\rt] $ is invariant under the abelian part of
   $(A4)$ viz.
   \be
   \de A_\m^{\al_{(0)}} = -\pa_\m\,\th^\al
   \ee
   Hence $ O_{(0)}\lt[A\rt] $ can be expressed in term of `abelian'
   Field strengths $ F^{( 0)}_{\m\n} = \pa_\m A^\al_\n -\pa_\n
   A^\al_\m $ and their derivatives,
   \be
  O_{(0)}[A]  = \, t^{\al_1\cdots \al_n}\,Y^{\al_1}_{0(1)}\cdots
  \, Y^{\al_n}_{0(n)}
  \ee
  where $ t^{\al_1\cdots\al_n} $ are constant and each $
  Y^{\al_k}_{0(k)} $ contains an arbitrary order of the derivative
  of $ F ^{(0)\al_k} $.
  \be
  Y^{\al_k}_{0(k)} =  \pa_{\m_1}\cdots \pa_{\m_q} \,
  F^{(0)\al_k}_{\n\la}
  \ee
   Further each $ O_{(p)}\lt[A\rt] $ must be separately invariant
   under global transformations of the group
   \be
   \de A^\al_\m = g f^{\al\bt\ga} A^\bt_\m \,\th^\ga; \,\,\,\,\,\,
   \th^\ga = \mbox{ constant } \nonumber
   \ee
   Nothing that each of $ Y ^{\al_k}_{0(k)} $ transforms as a {\em
   global vector} under $G$, it follows that the invariance of $
   O_{(0)}\lt[A\rt] $ under global transformations of $G$ require
   that $ t^{\al_1\cdots\al_n} $ is a group tensor.

  Now we construct
  \be
   Y^{\al_k}_{(k)}  = D^{\al_k \bt_1}_{\m_1}\,\,
   D^{\bt_1\bt_2}_{\m_2}\cdots D^{\bt_{q-1}\bt_q}_{\m_q} \,
   F^{\bt_q}_{\n\la}
   \ee
   Note that $Y_{0(k)}^{\al_k} = Y_{(k)}^{\al_k}\lt.\rt|_{g=0} $
   and

  \be
   \tilde{O}\lt[A\rt] = t^{\al_1\cdots\al_n}\,
   Y^{\al_1}_{(1)}\cdots Y^{\al_n}_{(n)}
   \ee
   Evidently $\tilde{O}\lt[A\rt] $ is a gauge invariant operator.
   Now consider $ O-\tilde{O}$. It is a gauge invariant operator of
   the same dimension $D$ and no term  in it of $0(g^0)$. It has
   the expansion
   \be
   O-\tilde{O} =\sum^D_{p=1}\, g^p\, O^\pr_{(p)}[A]
   \ee
   Now, we can repeat the discussion for the coefficient of lowest
   power of $g$ viz. $O^\pr_{(1)}\lt[A\rt] $. We then find a gauge
   invariant operator $\tilde{\tilde{O}} $ of the form of $(A10)$
   such that
   \be
   O-\tilde{O}-\tilde{\tilde{O}} =
   \sum^D_{p=2}\, g^p\, O^{\pr\pr}_{(p)}\lt[A\rt]
   \ee

   This process can be continued. It is guaranteed to end since the
   highest power of $g$ is finite $(D)$. Then
   \be
   O =\tilde{O} + \tilde{\tilde{O}} + \cdots
   \ee
   is of the form $(A10)$. Hence $(A1)$ is verified.

   Hence we can choose the basis for gauge invariant operators in
   which every term is of the form
   \be
  O = \, t^{\al_1\cdots \al_n}\,X^{\al_1}_{(1)}\cdots \, X^{\al_n}_{(n)}
  \ee

\se{}

We have noted in appendix $A$ that a typical gauge invariant
operator generalized to a 6-dimensional superspace contains
a chain of the form
\be
X^\al = D^{\al\bt}_i \cdots D^{\ga\de}_j\cdots F^\eta_{kl}
\ee
where the indices $i\cdots j\cdots kl $ are free to go from 0 to 5
 and when the group indices have been successively contracted. Now
 for a given chain all the indices $i\cdots j\cdots kl$ may be
 free(they may be contracted elsewhere in the expression  for $O$)
 or
 some of them may be contracted. Thus the chain $ X^\al $ may
 contain terms in which all of $i\cdots j\cdots kl $ take values
 between 0 to 3 and it may contain terms in which at least one of
 these is 4 and/or 5. In the latter group there terms in which 4
 and/or 5 appear {\it only } in $ F$ but not on any $D$ and finally
 there are terms in which at least one $D$ carries an index 4 or 5.
 It is this last type of terms that we focus our attention on. All
 the rest of the terms in $X$ are such that 4 and/or 5 appears only
 in $F$. We shall now show that the last type of terms can, by use
 of Jacobi identities, be cast as a sum over terms each of which has
 $ D_i $'s with $i$ going  only from 0 to 3; and 4 and/or 5 appears
 possibly only on $F$'s (or $c_5$). This would then show that the
 entire  $X$ has this property.

 For simplicity, first consider a simple example : Let $X$ contains
 only one $D$ with index (say) 4 and other $D$'s have only some
 space time index $ (\m = 0,1,2,3) $
 \be
X^\al = D^{\al\bt}_\m \cdots D^{\ga\de}_4 D^{\de\eta}_\n\cdots
F^\xi_{\sigma\la}
\ee
 We can now commute $D_4$ across $D_\ga $ and all the rest of further $D$'s
 by using (C2b) of appendix $C$. Finally when $D_4$ hits
 $F_{\sigma\la}$ we use (C1b) of appendix $C$ to express
 \bey
 D_4F_{\sigma\la} = -D_\la F_{4\sigma}-D_\sigma F_{\la 4} \nonumber
 \eey
 In each term in $X^\al$, now, the index 4 always appears on an
 $F$ and all $D$'s are with Lorentz indices.

 Now consider the case when the indices  on the last $F$ contain a
 4 and/or 5. All discussion is the same as before except when $D_4$
 hits $F$, we use one of the following ( see appendix $C$)
 \bey
 D_4 F_{4\m}&& = -\f{1}{2} D_\m F_{44} \nonumber \\
 D_4 F_{5\m} &&=-D_5 F_{4\m} - D_\m F_{45} = -gf c_5 F_{4\m} -D_\m
 F_{45} \lt(\mbox{ at } c_4 =0 = c_{4,\th}\rt) \nonumber \\
 D_4 F_{44} && =0 \nonumber \\
 D_4 F_{45} &&=-\f{1}{2} f c_5 F_{44} \lt(\mbox{ at }
 c_4=0=c_{4,\th}\rt) \nonumber \\
 D_4 F_{55} && =-2D_5F_{45} = 2f c_5 F_{45} \lt(\mbox{ at }
 c_4=0=c_{4,\th}\rt)
 \eey
 [We note  that in the evaluation in Sec III, we need to consider
 $\r{O}$ only at $ c_4 =0=c_{4,\th} $ ]
 proving the necessary result.

 Now consider a somewhat  more complicated case. Let there be two
 $D_4$ in $ X^\al$. We pass through first $D_4$ until it hits
 second; then use $ D_4D_4 =\f{g}{2} f F_{44} $. Next consider
  the case with three $D_4 $'s .
  \bey
X^\al = D^{\al\bt}_\m \cdots D^{\ga\de}_4\cdots D^{\eta\xi}_4
D^{\xi\ka}_\n\cdots D_4^{ \la\sigma} \cdots F^\tau_{ij} \nonumber
\eey
Here we need only exhibit how to pass third $ D_4^{\ga\de}$  after
second $ D_4^{\eta\xi}$ has been passed through.
\be
D_4^{\eta\xi}D_\n^{\xi\ka} = D_\n^{\eta\xi}D_4^{\xi\ka} + g
f^{\eta\ka\tau} F^\tau_{4\n}
\ee
This generates a term like $ F^{\eta\ka\tau} F^\tau_{4\n} $ through
which the third $ D_4 $ must be passed. This is done using the
identity
\be
D_4^{\ze\eta} f^{\eta\ka\tau} F^\tau_{4\n} =
f^{\ze\ka\tau}\lt(D_4^{\tau\eta} F^\eta_{4\n}\rt) - f^{\ze\eta\tau}
F^\tau_{4\n} D_4^{\eta\ka} = f^{\ze\ka\tau}\lt(-\f{1}{2}\rt)\lt(D_\m
F_{44}\rt) - f^{\ze\eta\tau}F^\tau_{4\n}D_4^{\eta\ka}
\ee
In the final term when all $ D$'s hit each other we use $D_4D_4D_4
=0$ Eq.(C1i). This process can be evidently generalized to more
than three $ D_4 $'s.

An exactly identical procedure follows when there are only $D_5$'s
(any number of them ) in $ X^\al$. Finally, we must deal with the
mixed case when there are both $D_4$'s and $D_5$'s. First consider
the simplest version of these
\bey
X^\al = D^{\al\bt}_\m \cdots D^{\eta\xi}_5\cdots D_4^{\la\sigma}\cdots
F^\ze_{ij} \nonumber
\eey
 The general procedure is similar to the previous cases and uses in
 particular modification of (B4). Ultimately, one ends up with

 \bey
 D_5 D_4 F_{ij} =\lt[ D_5D_4 -D_4 D_5\rt]- gfF_{45} F_{ij}
 \nonumber
 \eey
Now, it is easily shown using (C1) of appendix C that the term $
\lt(D_5D_4-D_4D_5\rt) F_{ij}$ case by case contains only $F$ 's and
$D_\m$'s but no $D_4$ or $D_5$ in the final result.
$\lt(c_4=0=c_{4,\th} \hbox{ has been emploed }\rt)$

As similar procedure can be given for more $D_5$ 's and/or more $ D_4$'s
Now,  what holds for each $X^\al$ hold for entire operator $O$ of
(B1). Thus $O$ can be written entirely
with out the use of $D_4 $ and $D_5$ anywhere ( ie in term  of
$ F_{ij},D_\m, c_5 $ only)

Now we consider the dependence of $ \r{O}_{\m_1\cdots\m_n}$  on
$A_{\n,\la} $. Firstly $ A_{\n,\la} $ is present only in some
$F_{4,\n}$ in $\r{O}$ (and not say $ D_4 $ acting on $ A_\n$
) because $O$ can be written with out  $D_4$. The subscript
4 must be contracted somewhere to 5 as $\r{O}_{\m_1\cdots\m_n} $
has no free subscript 4. Now consider the chain $ X $ which has
this index 5. From the above discussion ( with $ 4\leftrightarrow
5 $ everywhere ) it is clear that each such terms contains either
$F_{5\la} $ or $ F_{54}$ or $F_{55}$ a,
all of which vanish by equation of motion. The same applies to
$F_{44} $. This fact has been used in see III to simplify
integration over $A_{\m,\la},c_{4,\la} $ etc. in presence of an
operator.
Thus we have the result.

All terms in $\r{O}_{\m_1\cdots\m_n} $ depending on $ F_{4\m}$ and
$F_{44}$ vanishes by equation of motion. Those depending on
$F_{5\m}, F_{45}, F_{55} $ also vanishes because these quantities
themselves vanish by equation of motion.

\se{}

\vspace{.2in}

\sse{JACOBI IDENTITIES}
\begin{mathletters}
\bey
\label{eq:all}
D_\m F_{\n\la} +D_\n F_{\la\m} +D_\la F_{\m\n} = 0 \label{eq:a}\\
D_4 F_{\m\n} +D_\m F_{\n 4} + D_\n F_{4\m} =0   \label{eq:b}\\
D_5 F_{\m\n} +D_\m F_{\n 5} + D_\n F_{5\m} =0 \label{eq:c}\\
D_4 F_{5\m} -D_5 F_{\m 4} + D_\m F_{45} =0\label{eq:d} \\
2D_4 F_{4\m} + D_\m F_{44} = 0\label{eq:e} \\
2D_5 F_{5\m} + D_\m F_{55} = 0 \label{eq:f}\\
D_5 F_{44} + 2 D_4 F_{45} = 0\label{eq:g} \\
D_4 F_{55} + 2 D_5 F_{45} = 0 \label{eq:h}\\
D_4 F_{44} =0\label{eq:i} \\
D_5 F_{55} =0 \label{eq:j}
\eey
\end{mathletters}

\sse{COMMUTATION /ANTI-COMMUTATION RELATION OF THE COVARIANT
DERIVATIVES}
\vspace{.2in}
\begin{mathletters}
\bey
{\lt[D_\m ,\, D_\n \rt]}^{\al\bt} &=& -gf^{\al\bt\ga} F^\ga_{\m\n} \\
{\lt[D_4 ,\, D_\m \rt]}^{\al\bt} &=& -gf^{\al\bt\ga} F^\ga_{4\m} \\
{\lt[D_5 ,\, D_\m \rt]}^{\al\bt} &=& -gf^{\al\bt\ga} F^\ga_{5\m} \\
{\lt\{D_4 ,\, D_5 \rt\}}^{\al\bt}& =&  -gf^{\al\bt\ga} F^{\ga}_{45}\\
{\lt\{D_4 ,\, D_4 \rt\}}^{\al\bt}& = & -gf^{\al\bt\ga} F^{\ga}_{44}
\mbox{;   ie. } D_4^{\alpha \ga}D_4^{\ga\bt} = -\f{g}{2}f^{\al\bt\ga}F^\ga_{44}
\\
{\lt\{D_5 ,\, D_5 \rt\}}^{\al\bt} &=&  -gf^{\al\bt\ga} F^{\ga}_{55}
\mbox{;   ie. } D_5^{\alpha\ga}D_5^{\ga\bt} = -\f{g}{2}f^{\al\bt\ga}F^\ga_{55}
\eey

\end{mathletters}

\newpage

 \begin{references}

 \bi{brs} C. Becchi , A. Rouet, R. Stora. Comm.Math
 Phys,{\bf{42}}, 217 (1975).
 \bi{zinn} J. Zinn-Justin Lectures at "International Summer
 Institute for Theoretical physics" Bonn 1974.
 \bi{lee} See, for example, B. W.Lee, in {\em{Methods in Field Theory,
 Les Houches,France,1975}} edited by R. Balian and J. Zinn-Justin
 (North-Holland, Amsterdam .)
 \bi{ss1} For example, see K. Fujikawa, Prog.
 Theor.Phys.{\bf 59}, 2045 (1978) ; S. Ferrara, O. Piguet, M.
 Schweda, Nucl.Phys.{\bf B 119}, 493 (1977); L. Bonora and M.
 Tonin, Phys Lett. {\bf 98 B},48 (1981) ; A. C. Hirshfeld and H.
 Leschke, Phys. Lett. {\bf 101 B}, 48 (1981).
 \bi{ss2} R. Delbourgo and P. D. Jarvis , J. Phys {\bf A 15} 611
 (1982) ; L. Baulieu and J. Thierry-Mieg, Nucl. Phys.{\bf B 197}, 477
 (1982)
 \bi{sdj} S. D. Joglekar (unpublished) ; Phys. Rev. D {\bf 43}, 1307
 (1991);{\bf 48}, 1878(E) (1993).
 \bi{sdj1} S. D. Joglekar,Phys. Rev. D{\bf 44}, 3879 (1991);
 {\bf 48}, 1879(E) (1993).
 \bi{osp} S. D. Joglekar and B. P. Mandal, Phys. Rev. D{\bf 49}, 5382
 (1994).
 \bi{ren} S. D. Joglekar and B. P. Mandal, Phys. Rev. D {\bf 49}
 5617 (1994).
 \bi{anti} S. D. Joglekar and B. P. Mandal IIT Kanpur, preprint
 (1994).
 \bi{gross} See, eg. D. Gross and F Wilczek Phys. Rev D {\bf 8},
 3633 (1973) ; D {\bf 9}, 980 (1974).
 \bi{ann} S. D. Joglekar and B. W. Lee Ann. Phys, {\bf 97},
 160 (1976); See also, M. Henneaux Phys. Letts B {\bf 313}, 35 (1993).

\end {references}

\end{document}